\documentclass[11pt]{article}

\usepackage{aas_macros,amsmath,amssymb,comment,cite,esint,graphicx,mathtools}
\usepackage[margin=.8in,letterpaper]{geometry}
\usepackage[colorlinks=true]{hyperref}
\usepackage[affil-it]{authblk}
\usepackage{subcaption}
\usepackage[utf8]{inputenc}
\usepackage{mathrsfs}
\usepackage{appendix}
\usepackage{amssymb}
\usepackage{float}
\usepackage{color}
\usepackage{cite}
\usepackage{hyperref}
\hypersetup{pageanchor=false}
\usepackage{indentfirst}
\usepackage{url}
\usepackage{float}
\usepackage{caption}
\usepackage[numbers,square,comma,sort&compress,merge]{natbib}
\usepackage{esint}
\usepackage{overpic}
\usepackage{graphicx}
\usepackage{epsf,amsmath,bbold,amsfonts,stmaryrd}
\usepackage{textcomp}
\usepackage{ulem}
\usepackage{tikz}

\numberwithin{equation}{section}
\let\originalleft\left
\let\originalright\right
\renewcommand{\left}{\mathopen{}\mathclose\bgroup\originalleft}
\renewcommand{\right}{\aftergroup\egroup\originalright}
\mathcode`\*="8000
{\catcode`\*=\active\gdef*{\mathclose{}\,\mathopen{}}}

\newcommand{\td}[1]{\tilde{#1}}

\newcommand{\be}{\begin{equation}}
\newcommand{\ee}{\end{equation}}
\newcommand{\bea}{\setlength\arraycolsep{2pt} \begin{eqnarray}}
\newcommand{\eea}{\end{eqnarray}}
\newcommand{\nn}{\nonumber}

\def\a{\alpha}
\def\b{\beta}

\def\d{\delta}
\def\D{\Delta}
\def\f{\frac}
\def\g{\gamma}

\def\lm{\lambda}

\def\m{\mu} 
\def\n{\nu} 
\def\nn{\nonumber}
\def\pl{\partial}
\def\p{\phi} 
\def\td{\tilde} 

\def\s{\sigma}

\def\t{\theta}

\def\ep{\epsilon}
\def\o{\omega}

\def\be{\begin{equation}}
\def\ee{\end{equation}}
\def\bag{\begin{aligned}}
\def\eag{\end{aligned}}
\def\bea{\begin{eqnarray}}
\def\eea{\end{eqnarray}}
\def\ba{\begin{array}}
\def\ea{\end{array}}

\def\bc{\begin{center}}
\def\ec{\end{center}}

\begin{document}
\title{Energy Extraction from a Kerr Black Hole via Magnetic Reconnection within the Plunging Region}

\author{
Bin Chen$^{1,2,3}$\thanks{E-mail: bchen01@pku.edu.cn}~, 
Yehui Hou$^{2}$\thanks{E-mail: yehuihou@pku.edu.cn}~, 
Junyi Li$^{2}$\thanks{E-mail:  2100011312@stu.pku.edu.cn}~, 
Ye Shen$^{2}$\thanks{E-mail: shenye199594@stu.pku.edu.cn}~}
\date{}

\maketitle
\vspace{-10mm}

\begin{center}
	{\it	
	$^1$ Department of Physics, School of Physical Science and Technology, Ningbo University, Ningbo, Zhejiang 315211, China\\\vspace{4mm}
	
		$^2$School of Physics, Peking University, No.5 Yiheyuan Rd, Beijing
		100871, P.R. China\\\vspace{4mm}
		
		$^3$Center for High Energy Physics, Peking University,
		No.5 Yiheyuan Rd, Beijing 100871, P. R. China\\\vspace{4mm}
		
	}
\end{center}

\vspace{8mm}

\begin{abstract}
	\vspace{5mm}
Magnetic reconnection within a highly magnetized plasma has been seen as a viable mechanism to extract the energy from a rotating black hole, as it can generate negative energy plasmoids in the ergoregion. For a typical accreting black hole, the ergoregion is filled with bulk plasma plunging from the innermost-stable-circular orbit (ISCO). In this study, we present an analytical study of the energy extraction via magnetic reconnection process in the plunging region. In contrast to the toroidal plasma, where the magnetic field cannot be derived from the MHD scheme, the magnetic field in the plunging plasma was determined by the ideal-MHD condition. We derive the global magnetic field structure in a fast reconnection model, and we read the expressions for the energies of plasmoids ejected from the reconnection region, for general stationary and axisymmetric spacetimes. Then, we demonstrate the behaviors of ejection energies varying with the reconnection locations in the Kerr spacetime, and identify the region where a negative-energy plasmoid can be produced. We find that for a certain magnetization there exists a critical value of the black hole spin, beyond which the energy extraction can occur, and the energy extraction is most efficient for the near-extreme black hole. Moreover, we study the conditions necessary for a plasmoid with positive energy to escape to the infinity, a crucial requirement for effective energy extractions. Considering the escaping conditions, we provide the parameter space in the radius-spin plane in which the energy extraction mechanism is effective. 
\end{abstract}

\maketitle

\newpage
\baselineskip 18pt

\section{Introduction}\label{sec1}
	
Energy extraction from a rotating black hole has been a long-standing topic of interest in astrophysics. It has been demonstrated that the rotational energy of a Kerr black hole can be extracted through classical processes without violating the black hole's second law, until the black hole loses its mass to the irreducible value \cite{Christodoulou:1970wf}. A prime example is the Penrose process, where a particle undergoes fission within the ergosphere, producing two particles. One particle, possessing negative energy as viewed from the infinity, falls into the black hole, while the other, carrying more energy than the original particle, escapes to a distance. Although theoretically plausible, this process suffers from a lack of initiation mechanisms \cite{Wald:1974kya}. Instead, the superradiance of a massive bosonic field has been viewed as a practical mechanism. In this scenario, a classical field forms a superradiant cloud with a specific spectrum around the black hole \cite{Brito:2015oca}. Another mechanism for extracting the energy of a rotating black hole involves magnetic fields. In the case of black holes surrounded by magnetized accretion flows, Blandford and Znajek \cite{Blandford:1977ds} discovered that a stationary electromagnetic field can extract the black hole's energy if the magnetic field lines intersect the event horizon and possess an angular velocity smaller than that of the horizon. The Blandford-Znajek (BZ) mechanism serves as a potent engine for generating relativistic jets, a claim supported by both numerical studies and observations  \cite{Komissarov:2005wj, Parfrey:2018dnc, Lu:2023bbn}.

Beyond the above mechanisms, it has been proposed theoretically that the energy extraction is viable via magnetic reconnections. Magnetized accretion flows around black holes can liberate electromagnetic energy in the regions where magnetic field lines are highly curved and reconnect to discharge their energy \cite{yamada2009, SP1, Petschek, rela-SP}. This energy release propels the plasma ejected from the reconnection region. If the released magnetic energy is sufficiently strong, the accelerated plasma can obtain sufficiently negative angular momentum and negative energy, as viewed from the infinity. Koide and Arai proposed the possibility of energy extraction from black holes through magnetic reconnection \cite{Koide:2008xr}, where they treated the ejected plasma as relativistic adiabatic incompressible ball, without delving into the detailed dynamics of the plasmoids. More recent studies has shown that the plasmoid instability within the reconnection layer and the generation of expelled plasmoids can lead to fast reconnection and fast Alfv{\'e}n outflows from these layers \cite{Sironi:2022hnw,French:2022zfv,Dong:2022crn,Comisso:2023ygd}. Subsequently, Comisso and Asenjo \cite{Comisso:2020ykg} investigated the energy extraction through the plasmoids in a fast reconnection model \cite{Liu2017}. They calculated the energy extraction efficiency and power, showing that such an energy extraction rate could be comparable to that of the BZ process, thereby implying its astrophysical significance. Subsequent studies have explored the energy extraction via magnetic reconnections in different types of spacetimes \cite{Khodadi:2022dff, Carleo:2022qlv, Wei:2022jbi, Liu:2022qnr, Wang:2022qmg, Li:2023nmy, Li:2023htz, Ye:2023xyv, Khodadi:2023juk, Shaymatov:2023dtt, Zhang:2024rvk}.

To date, most of studies on energy extraction via magnetic reconnections have been primarily focused on circular accretion flows. However, circular orbits within the ergosphere are largely unstable, since the ISCO of a Kerr black hole lies outside the ergosphere, as long as $a < 0.943M$. It is unlikely for the plasma to maintain a circular orbit within the ISCO, as it would escape under small perturbations. A more realistic scenario is that the accretion flows are circular outside the ISCO, while they plunge toward the event horizon inside the ISCO, adopting a spiral shape \cite{Mummery:2022ana, Liu:2023tcy}. Such a plunging region has been observed and highlighted in various astrophysics-related studies \cite{Machida:2002ub, Reynolds:2006uq, Zhu:2012vf, Wilkins:2020pgu, Hankla:2022iwk, Dong:2023bbd}. It is expected that in the plunging region, the behavior of the ejected plasmoids produced from magnetic reconnections could present significantly different features, given that the magnetic field structure and the ejection directions of the plasmoids are distinct from those in toroidal flows.

In this study, we present an analytical investigation of energy extraction via magnetic reconnection in the plunging region near a rotating black hole. Our analysis begins with a general stationary and axisymmetric spacetime, assuming that the accretion flow in the equatorial plane conforms to the Killing symmetry of the spacetime. One remarkable feature in our study is that the global magnetic field within the plunging plasma can be determined based on the calculation presented in \cite{Ruffini:1975ne}, in contrast to the situation in the toroidal plasma. For the detailed structure of the reconnection region, we utilize a fast reconnection model from \cite{Liu2017}, characterized by strong magnetic pressure. Subsequently, we derive a general expression for the energies of the plasmoids ejected from the reconnection region near the equatorial plane. Within this framework, we focus on the Kerr black hole and consider magnetic reconnection occurring in the plasma within the ISCO, with the streamlines following plunging geodesics. We explore how the energies of the ejected plasmoids vary with the positions of the reconnection region under different black hole spin and magnetization. Furthermore, we discuss the conditions necessary for a plasmoid with positive energy to escape to infinity, a crucial requirement for successful energy extraction.

The paper is organized as follows. In Sec.~\ref{sec:basic}, we discuss the general aspects of the analytical model.  We introduce the basic equations of the fluid model in Sec.~\ref{sec:plasma} and the model of fast reconnection in Sec.~\ref{sec:Rmodel}, present the global structure of the magnetic field  in Sec.~\ref{sec:mag_field}, and derive the general expression for the energy-at-infinity per enthalpy of the ejected plasmoids  in Sec.~\ref{sec:energy}. We apply the model to investigate the energy extraction in the plunging region of the Kerr spacetime in detail in Sec.~\ref{sec:EE_inspiral}, including the specifics of the magnetic field angle in Sec.~\ref{MFA},  the behaviors of the energy-at-infinity per enthalpy of the ejected plasmoids and local energy extraction efficiency in Sec.~\ref{Ee}, and the escaping conditions in Sec.~\ref{Ec}. We summarize our work in Sec.~\ref{sec:sum}. The unit used throughout the paper is $G = c = 1$.

\section{Magnetic reconnections: model and setup}
\label{sec:basic}
To begin with, let us revisit some basic concepts related to magnetic reconnections in a stationary, axisymmetric spacetime. The line element of the spacetime can be written in the 3+1 form as
\bea
    ds^2 = g_{\m\n} dx^{\m} dx^{\n} = -\a^2 dt^2 + \sum_{i = r, \t, \p}  \big(h_i dx^i - \a \b^i dt \big)^2 \, ,
\eea
where 
\be\label{alpha}
    \a = \sqrt{-g_{tt} + \f{g^2_{t\p}}{g_{\p\p}}} \, ,\quad  h_i = \sqrt{g_{ii}} \, , \quad  
    \b^i = \delta_{i \p}\f{\sqrt{g_{\p\p}}\, \o^{\p}}{\a} \, 
\ee
with $\o^{\p} = -g_{t\p}/g_{\p\p}$ being the angular velocity of frame dragging. The normal observers of the spacetime, often referred to as zero-angular momentum observers (ZAMOs), are defined via the normal tetrad
\bea
\hat{e}^{\, \m}_{(t)} = \f{1}{\a} (\pl_t^{\, \m} + \o^{\p} \pl_{\p}^{\, \m}) \, ,\quad  \hat{e}^{\, \m}_{(r)} = \f{1}{h_r} \pl_r^{\, \m} \, ,\quad     \hat{e}^{\, \m}_{(\t)} = \f{1}{h_\t} \pl_\t^{\, \m} \, ,\quad   \hat{e}^{\, \m}_{(\p)} = \f{1}{h_{\p}} \pl_{\p}^{\, \m} \,.
\eea 

\subsection{The bulk plasma}
\label{sec:plasma}

Strictly speaking, the horizon-scale plasma is nearly collisionless, with the electron free path possibly comparable to the gravitational radius. The intrinsic properties, such as the energy distributions of thermal electrons in both the ejected plasmoids and the bulk plasma, should be described within the framework of Vlasov gas dynamics. However, for simplicity, our current study employs the perfect fluid approximation. The stress-energy tensor of the bulk plasma is given by
\be
    T^{\m\n} =\o\, U^{\m}U^{\n} + p\, g^{\m\n}\, ,
\ee
where $p$, $\o$ are the fluid's proper pressure and enthalpy density, while $U^{\m}$ is the 4-velocity of the streamlines of the fluid. The energy conservation law for the fluid, $\nabla_{\m} T^{\m}_{\,\,\,\,t} = 0$, can be rewritten as
\bea
    \pl_t e + \f{1}{h_1h_2h_3} \pl_i ( h_1h_2h_3 S^i) = 0 \, ,
    \label{energy}
\eea
where $e =- \a T^t_{\, \, \, t}$ is called the energy-at-infinity density, and $S^i = -\a T^i_{\, \, \, t}$ is the flux density. In our work, we are interested in a stationary, axisymmetric and geometrically thin fluid configuration. Thus, all the physical quantities are defined over the equatorial plane, neglecting their $\theta$-dependence. The flow is considered to be dominated by the gravity and flow along timelike geodesics, such that $U_\m dx^\m= -E dt +U_r(r) dr + L d\p$, with $E,L$ being the conserved quantities along the streamline.

As our focus is on the occurrences within the fluid during magnetic reconnection, it is necessary to quantify the processes in the fluid's rest frame. To do this, we firstly project the fluid 4-velocity into the normal tetrad,
\bea
    \hat{U}^{(a)} = U^\m \hat{e}_{\, \m}^{(a)} = \hat{\g}_s\left\{1,\hat{v}^{(r)}_s,0,\hat{v}^{(\p)}_s \right\} = \left\{ \f{E - \o^{\phi}L}{\a}, \sqrt{g_{rr}} \, U^r, 0,  \f{L}{\sqrt{g_{\phi\phi}}} \right\} \, .
\eea
Then, the tetrad for the fluid's rest frame can be defined through the normal tetrad as
\bea
    &&e_{[0]} = \hat{\g}_s \left[\hat{e}_{(t)}+\hat{v}^{(r)}_s \hat{e}_{(r)}+\hat{v}^{(\p)}_s \hat{e}_{(\p)}\right]  \, , \nn \\
    &&e_{[1]} = \f{1}{\hat{v}_s} \left[\hat{v}^{(\p)}_s \hat{e}_{(r)}-\hat{v}^{(r)}_s \hat{e}_{(\p)}\right]  \, , \quad  e_{[2]}  = \hat{e}_{(\t)}  \, , \nn \\
    &&e_{[3]} = \hat{\g}_s \left[\hat{v}_s \hat{e}_{(t)}+\f{\hat{v}^{(r)}_s}{\hat{v}_s} \hat{e}_{(r)}+\f{\hat{v}^{(\p)}_s}{\hat{v}_s} \hat{e}_{(\p)}\right]  \, ,
    \label{eq:tetrad}
\eea
where $\hat{v}_s = \sqrt{\left(\hat{v}^{(r)}_s\right)^2+\left(\hat{v}^{(\p)}_s\right)^2}$.
Clearly, for a normal observer, $e_{[1]}, e_{[3]}$ is orthogonal and parallel to the fluid velocity, respectively.

\subsection{The magnetic-reconnection model}
\label{sec:Rmodel}

We now turn to a brief review of the magnetic-reconnection models. The Sweet-Parker model \cite{SP1,SP2}, a typical steady-state model, has been extensively applied in the related studies. However, this model is characterized by a low reconnection rate, which fails to account for numerous astrophysical processes such as solar flares and substorms \cite{yamada2009, substorm}. The Petschek model \cite{Petschek} offers a remarkably higher reconnection rate, but its self-consistency has been questioned \cite{anti-Petschek1,anti-Petschek2}. 

\begin{figure}[h!]
	\centering
	{\includegraphics[width=\textwidth]{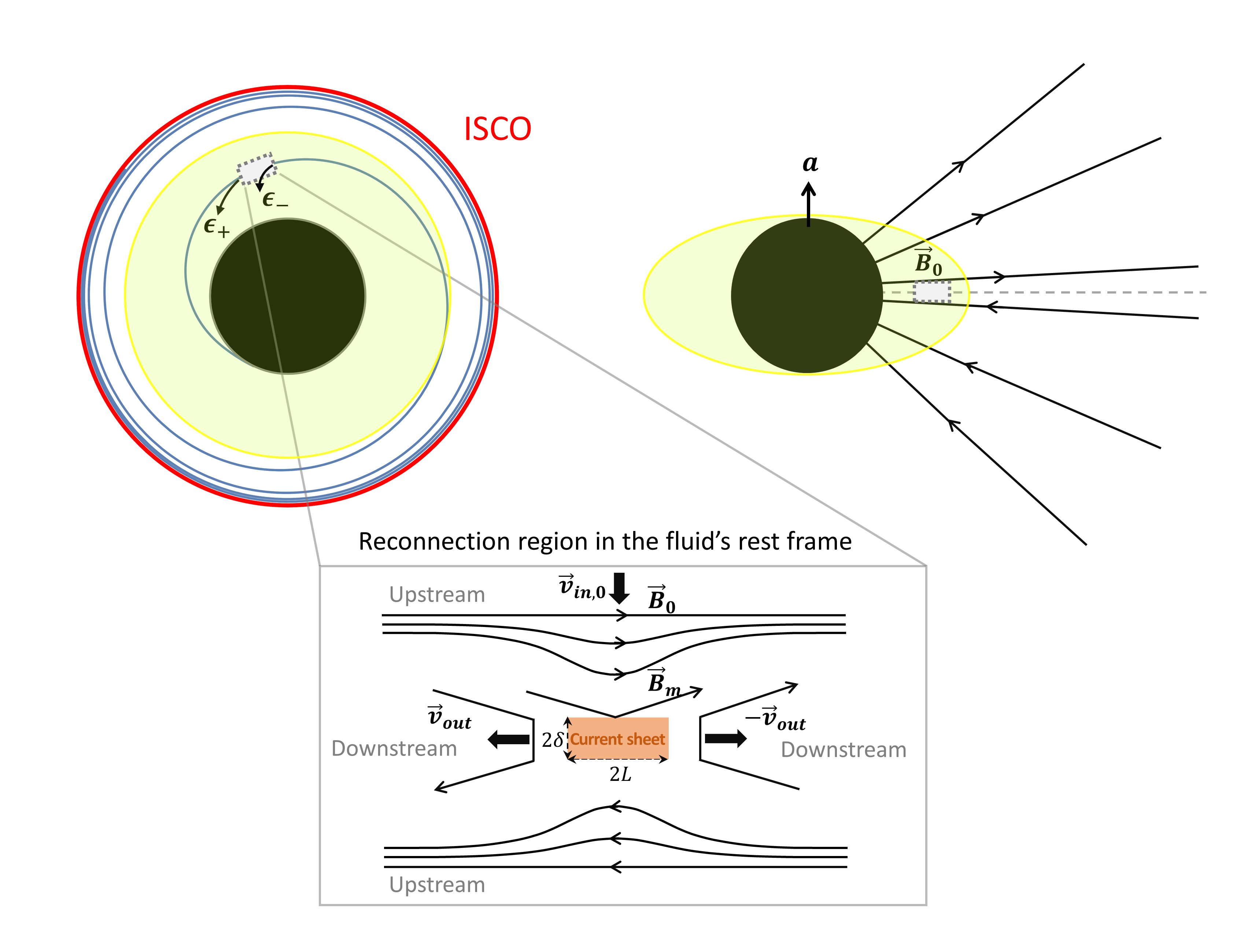}}
	\caption{Schematic diagram of the magnetofluid at various scales. Top left: the plunging plasma viewed from above. The red circle represents the ISCO, with the blue spiral curve depicting a streamline of the bulk plasma. The light yellow region represents the ergo-region. Two plasmoids are ejected from the reconnection region (small gray square). Top right: the global magnetic field viewed laterally, where the black arrowed lines represent the magnetic field lines. Bottom: a closer look at the magnetic structure on the local scale, where $\vec{B}_0$ denotes the global magnetic field, and $\vec{B}_m$ denotes the magnetic field in the current sheet (small orange square); $\vec{v}_{in,0}$ is the local inflow speed of the plasma, and $\vec{v}_{out}$ denotes the velocities of plasmoids ejected from the current sheet.}	
	\label{illu}
\end{figure}

In this study, we consider the fast reconnection model within the MHD framework \cite{Liu2017}, which estimates a peak reconnection rate to be about $0.1$. This model assumes that the thermal pressure is negligible relative to the magnetic pressure, an assumption that is generally applicable in highly magnetized accretion systems, particularly in magnetically arrested disks \cite{MAD,Tchekho2011}. The scheme is illustrated in Fig.~\ref{illu}. The top two panels illustrate the plunging bulk plasma and the global magnetic field near the equatorial plane, both of which exist on a global scale defined by the gravitational radius, $r_g = GM/c^2$. Within this scale, the magnetic field lines reverse directions upon crossing the equatorial plane, as shown in the top right panel of Fig.~\ref{illu}. Compared to the gravitational radius, the magnetic reconnection takes place within a much small region \cite{Priest1986}, proximal to the dominant reconnection $X$-point \cite{Titov1997}. Therefore, we consider the location of the reconnection region as the $X$-point on the global scale.

The bottom panel of Fig.\ref{illu} provides a detailed illustration of the magnetic field structure around the $X$-point at a local scale. This smaller MHD scale is where the physical quantities of the bulk plasma achieve uniformity. The local scale is significantly smaller than the gravitational radius, yet larger than the mean free path of ions, ensuring the feasibility of the MHD scheme. Two sets of anti-parallel magnetic field lines upstream, represented by $\vec{B}_0$, are gradually curved towards each other due to spontaneous perturbations. This curvature reduces the local magnetic pressure below its original value, and the resulting pressure differential continues to push the anti-parallel magnetic lines towards each other until they meet at the diffusion region (the current sheet). This is where reconnection occurs and two resultant plasmoids are ejected downstream, moving along the direction of the upstream magnetic field line $\vec{B}_0$. The current sheet is depicted by an orange square in the bottom panel of Fig.\ref{illu}, with a size comparable to the mean free path of ions in the fast reconnection model \cite{yamada2009}. We denote the width and length of the current sheet as $2\d$ and $2L$ respectively, and define a geometric index as
\bea\label{eq:geo_index}
    {\sf g}=\frac{\delta}{L} \, .
\eea
Assuming the plasmoids are not decelerated after leaving the current sheet, we can estimate the magnitude of their ejection velocities using the MHD equilibrium, which yields
\bea
    v_{out} \simeq \sqrt{\frac{\left(1-{\sf g}^2\right)\sigma_m}{1+\left(1-{\sf g}^2\right)\sigma_m}} \, ,
    \label{eq:vout}
\eea 
where $\sigma_m=B_{m}^2/\omega$ represents the magnetization in the current sheet. As ${\sf g}$ approaches zero, the ejection velocity reduces to that of the Sweet-Parker model \cite{rela-SP}. Moreover, the limit $\lim\limits_{{\sf g}\to 1}v_{out}=0$ imposes a restriction on the geometric index. This restriction implies that, if reconnection occurs, the magnetic pressure differential between $\vec{B}_0$ and $\vec{B}_m$ should be capable of overcoming the magnetic tension, which increases when the magnetic field line curves. The ratio of magnetization in the diffusion region to the upstream magnetization reads
\bea
    \frac{\sigma_m}{\sigma_0} \simeq \frac{B_{m}^2}{B_{0}^2} \simeq \left(\frac{1-{\sf g}^2}{1+{\sf g}^2}\right)^2 \, ,
    \label{eq:mto0}
\eea
given by the local-scale-MHD equilibrium. Combining Eq.~\eqref{eq:mto0} with Eq.~\eqref{eq:vout}, we get the ejection velocity expressed by the upstream magnetization,
\bea
    v_{out} \simeq \sqrt{\frac{\left(1-{\sf g}^2\right)^3 \sigma_0}
                {\left(1+{\sf g}^2\right)^2+\left(1-{\sf g}^2\right)^3 \sigma_0}} \, .
    \label{eq:vout0}
\eea
We further introduce a local reconnection rate, denoted as $R_0$, to describe the generation rate of magnetic reconnection. This rate is defined as the ratio of the local inflow speed ($v_{in,0}$ in Fig.~\ref{illu}) to the Alfv{\'e}n velocity. The behavior of $R_0$ varies with different magnetizations. In the case of low local magnetization ($\sigma_0\ll 1$), the local-reconnection rate is solely dependent on the geometric index,
\bea
    R_{0}\simeq {\sf g}\sqrt{1-{\sf g}^2}\left(\frac{1-{\sf g}^2}{1+{\sf g}^2}\right)^2 \, ,
    \label{eq:RLLM}
\eea
which peaks at ${\sf g}\simeq 0.31$. In the case of high local-magnetization limit ($\sigma_0\gg 1$), we have
\bea
    R_{0}\simeq {\sf g}\frac{1-{\sf g}^2}{1+{\sf g}^2}
                    \sqrt{\frac{\left(1-{\sf g}^2\right)^3 \sigma_0}{\left(1+{\sf g}^2\right)^2+\left(1-{\sf g}^2\right)^3 \sigma_0}} \, ,
    \label{eq:RHLM}
\eea
which peaks at ${\sf g}\simeq 0.49$. In subsequent discussions, we will consider the high local-magnetization limit and assume that magnetic reconnection always occurs at the maximum local-reconnection rate. Therefore, we set the geometric index to be ${\sf g}=0.49$. 

\subsection{The global magnetic field structure}
\label{sec:mag_field}

In this section, we provide an explicit expression for the global magnetic field near the equatorial plane, which corresponds to the upstream magnetic field at the local scale. We assume that the magnetic field lines above and below the equatorial plane are identical, but with opposite directions. Therefore, in the subsequent discussion, we will focus solely on the magnetic field structure above the equatorial plane.

The conductivity of the plasma is nearly zero outside the current sheet, which means that the ideal-MHD condition, $F_0^{\m\n}U_{\m} = 0$, holds, where $F_0^{\m\n}$ is the electromagnetic tensor. For a stationary, axisymmetric fluid configuration composed of zero-conductivity plasma, the global magnetic field is determined by the ideal-MHD condition \cite{Ruffini:1975ne}, and takes the form of 
\bea
    B_0^t = -\f{\Psi}{\sqrt{-g}} \bigg( U_r + \f{U^{\phi}}{U^r}L \bigg) \, , \quad B_0^r = -\f{\Psi}{\sqrt{-g}} E \, , \quad B_0^\phi = -\f{\Psi}{\sqrt{-g}} \f{U^{\phi}}{U^r} E \, .
    \label{B}
\eea
Here $g$ is the determinant of the metric, $\Psi$ denotes the overall strength and keeps constant along the streamline.\footnote{For our propose, we require $\Psi \rightarrow -\Psi$ under $\t \rightarrow \pi - \t$. As the region for magnetic reconnection is geometrically thin around the equatorial plane, we can regard $\Psi$ as a constant on either side of the equatorial plane.} Note that Eq.~\eqref{B} only works for non-circular streamlines. For circular fluid with $U^r = 0$, the above expression diverges, which means that the magnetic field in this case cannot be obtained directly from the ideal-MHD condition. 

In the fluid's rest frame, the magnetic field components take the forms of \cite{Hou:2023bep}
\bea
    B_0^{[1]} = \f{\Psi}{\sqrt{-g} \,\hat{U}} \sqrt{g_{rr}g_{\p\p}}\,\o^{\p}\, , \quad  B_0^{[3]} =  \f{\Psi}{\sqrt{-g} \,\hat{U} U^r}  (\hat{\g}_s E-\a) \,,
    \label{B1}
\eea
where $\hat{U} = \sqrt{\left(\hat{U}^{(r)}\right)^2+\left(\hat{U}^{(\p)}\right)^2}$. It is essential to know the direction of the global magnetic field in the fluid's rest frame, along which the plasmoids are ejected. To quantify this direction, we define the magnetic field angle as the angle formed between the magnetic field line and $e_{[3]} $,
\bea
    \xi_B  = \arctan{ \f{B_0^{[1]}}{B_0^{[3]}} } = \arctan{\bigg(\f{\sqrt{g_{rr}g_{\phi\phi}}\,\o^{\phi}U^r}{\hat{ \g}_s E -\a} \bigg)}\, .
    \label{angle}
\eea
For the fluid accreted by the black hole, $U^r < 0$. Then, from $\hat{ \g}_s \ge 1$ one can deduce $E \ge \a + \o^{\p}L \ge \a$ for a prograde flow. Thus, it is easy to see from Eq.~\eqref{angle} that $\xi_B$ is always negative for an accreted, prograde flow. 

Furthermore, it is important to note that the derivation of the global magnetic field mentioned earlier does not explicitly rely on the fluid assumption. What we established is a stationary, axisymmetric, magnetized plasma system characterized by a bulk velocity $U^{\m}$ at each position. The plasma exhibits sufficiently high conductivity, such that in its rest frame the electrons promptly react and reorganize to neutralize any electric field $E^{\m} = F^{\m\n}U_{\n}$ that might exist. The timescale associated with this conductivity is shorter than the plunging timescale, ensuring a consistent $E^{\m} = 0$. These conditions allow for the derivation of $B^\m(r, \theta)$. Further details can be found in \cite{Ruffini:1975ne, Hou:2023bep}. 

\subsection{Energies of ejected plasmoids}\label{sec:energy}

In the magnetic reconnection process, the ejection velocities of plasmoids are aligned with the magnetic field line at the $X$-point. Projected to the fluid's rest frame, the four-velocities of the ejected plasmoids can be written as
\bea
    u^{\m}\pl_{\m} =  \g_{out}\big[ e^{[0]}  \pm v_{out} \big(\cos{\xi_B} e^{[3]} + \sin{\xi_B} e^{[1]} \big)\, \big] \, 
    \label{4v}
\eea 
with $\g_{out}$ being the Lorentz factor measured in the fluid's rest frame. The ''$\pm$'' represents two plasmoids towards opposite directions. 

During the reconnection process, most of the magnetic energy is converted to the plasma kinetic energy in the current sheet. Thus, within an plasmoid, we can neglect the magnetic energy and only consider the fluid part. On one hand, we take the incompressible sphere approximation, under which the proper thermal quantities of plasma keep constant during the reconnection process \cite{Koide:2008xr}. Thus, the proper enthalpy density in the plasmoid is equal to that of the bulk plasma at the $X$-point, and the enthalpy reads $\D H = \o \D V$, assuming that the plasmoid has a small proper volume $\D V$. On the other hand, the fluid energy-at-infinity density is defined in the normal tetrad, taken to be $e =- \a T^t_{\, \, \, t} = -\a (\o u^t u_t + p)$, as has been introduced in Eq.~\eqref{energy}.  The volume of the plasmoid measured by the normal tetrad is  $\D \hat{V} =  \D V/\hat{u}^{(t)} =   \D V/( \a u^t)$, and the energy is $\D E = e \D \hat{V}$. Based on these, we can define the energy-at-infinity per enthalpy for the ejected plasmoids at the $X$-point as 
\bea
 \ep = \f{\D E}{\D H} = \f{e \D \hat{V}}{\o \D V} = \f{e}{\a u^t \o} =  -u_t - \f{\td{p}}{u^t} \, ,
 \eea
where $\td{p} = p/\o$ is the pressure per enthalpy. In what follows, we use $\ep$, as a widely-accepted choice in previous works, to represent the magnitude of energy obtained by the plasmoid from magnetic reconnection in the following discussion. Reading $u^t, u_t$ from Eq.~\eqref{4v},  we have
\begin{align}\label{ep}
\ep_{\pm} = &\, \, \a \hat{\g}_s \g_{out} \bigg[  \big(1+\b^{\p}\hat{v}^{(\p)}_s\big) \pm  v_{out}\bigg(\hat{v}_s+ \b^{\p}\f{\hat{v}^{(\p)}_s}{\hat{v}_s}\bigg)\cos{\xi_B} \mp v_{out}\b^{\p}\f{\hat{v}^{(r)}_s}{\hat{\g}_s \hat{v}_s} \sin{\xi_B}   \bigg] \nn \\
&- \f{\a \, \td{p}}{\hat{\g}_s \g_{out}\big(1\pm \hat{v}_s v_{out} \cos{\xi_B}\big)} \, .
\end{align}
Note that Eq.~\eqref{ep} is analogous to Equation (28) in \cite{Comisso:2020ykg}. The above discussion is quite general and applicable to any stationary and axisymmetric spacetime. According to Eq.~\eqref{ep}, it is clear that the influence of the spacetime is directly encoded in $\a$ and $\b^{\p}$, while indirectly encoded in the fluid velocity and the magnetic field angle. The value of $v_{out}$ depends on the reconnection model, which has no association with the background spacetime.

\section{Energy extraction from a Kerr geometry}
\label{sec:EE_inspiral}

In this section, we focus on energy extraction from a Kerr black hole via magnetic reconnections within the plunging region. We consider the bulk plasma that is perturbed and plunges into the black hole from ISCO. The radii of ISCOs are
\bea
&&r_{ms} = 3 + Z_2 \mp \sqrt{(3-Z_1)(3+Z_1+2 Z_2)} \, , \nn \\
&&Z_1 = 1 + (1-a^2)^{1/3} \big[(1+a)^{1/3}+ (1-a)^{1/3}\big]\, , \quad Z_2 = \sqrt{3a^2+Z_1^2} \, ,
\eea 
where $a$ denotes the black hole spin, $\mp$ represents the prograde/retrograde orbit. Here and thereafter, we set the ADM mass of the Kerr black hole to be $M = 1$, without loss of generality. The conserved energy and angular momentum of the geodesics are inherited from the ISCOs, taking the form of
\be\label{iscoEL}
    E_{ms} = \f{r^{3/2}_{ms} - 2\sqrt{r_{ms}} \pm a}{\sqrt{r^3_{ms} - 3 r^2_{ms} + 2a r^{3/2}_{ms}}} \, ,
   \quad L_{ms} = \pm \f{r^{2}_{ms} \mp 2a\sqrt{r_{ms}} + a^2}{\sqrt{r^3_{ms} - 3 r^2_{ms} + 2a r^{3/2}_{ms}}} \, .
\ee
In what follows we only consider the prograde case.\footnote{For the retrograde case, we confirm that the plasmoids have $\epsilon_{\pm} > 0$, even under strong magnetizations. Thus, energy extractions via magnetic reconnection cannot occur within the retrograde plasma.} In the Kerr geometry, the ISCO is the triple root of the effective potential for radial motion, leading to a simple form of the radial velocity of the plunging geodesics \cite{chandrasekhar1998mathematical},
\bea
    U^r = -\sqrt{\f{2}{3\,r_{ms}}} \bigg( \f{r_{ms}}{r} -1\bigg)^{\f{3}{2}} \, .
    \label{iscour}
\eea

\subsection{The magnetic field angle}\label{MFA}

\begin{figure}[h!]
	\centering
	{\includegraphics[width=0.48\textwidth]{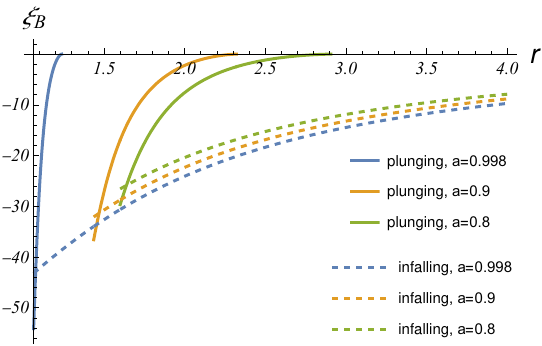}}\,
	{\includegraphics[width=0.48\textwidth]{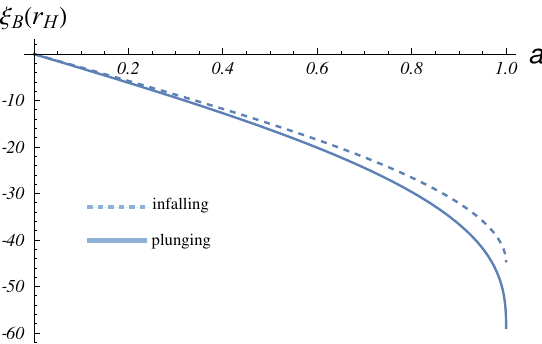}}
	\caption{Behaviors of the magnetic field angle in the bulk plasma plunging from the ISCO (Solid lines) in Kerr geometry. We also plot the results for the infalling plasma with $E = 1, L = 0$ as a comparision (Dashed lines) . Left: $\xi_B$ as a function of $r$ for different $a$ (0.8, 0.9, 0.998). Right: The value of $\xi_B$ at the event horizon $r_H = 1+ \sqrt{1-a^2}$, as a function of $a$. }
	\label{xiB}
\end{figure}

In Fig.~\ref{xiB} we present the plot of the magnetic field angle, as defined in Eq.~\eqref{angle}. As we can see from the left panel, $\xi_B$ is consistently negative, starting from zero at the ISCO and decreasing with the decrease of the radius. As the misalignment between the magnetic field lines and streamlines is a consequence of the frame dragging effect, it intensifies as one gets closer to the horizon of the black hole. Moreover, when the fluid is in close proximity to the ISCO, its motion is predominantly in the $\phi$-direction, hence confining the magnetic field to the direction of $e_{[3]}$.  

As shown on the right panel in Fig.~\ref{xiB}, the magnetic field angle at the horizon decreases fastly with the increasing spin, approaching a minimum value of $-60^{\circ}$ as $a \rightarrow 1$. To make a clear analysis for the near-extreme case, we define $a = \sqrt{1-\lm^2}$, $\lm \ll 1$, and expand the physical quantities to the powers of $\lm$ to see their asymptotic behaviors. The event horizon takes $r_H = 1+ \sqrt{1-a^2}=  1+ \lm$. The quantities related to the ISCO can be expanded, to $ \mathcal{O}(\lm^{2/3}) $,
\begin{align}
r_{ms} \approx 1 + 2^{1/3} \lm^{2/3} \, , \quad
E_{ms} \approx \f{1}{\sqrt{3}} + \f{2^{1/3}}{\sqrt{3}} \lm^{2/3} \, , \quad L_{ms} \approx \f{2}{\sqrt{3}}  + \f{2^{4/3}}{\sqrt{3}} \lm^{2/3} \, .
\end{align}
The near-horizon region can be recovered by taking a coordinate transformation, $r = \lm R +1$, $R \ge 1$ \cite{Bardeen:1999px}. In this region, the magnetic field components can be expanded as
\begin{align}
B^{[1]} & \approx \f{\Psi}{\lm} \sqrt{\f{3}{R^2+3}}+ \mathcal{O}(\lm^{-2/3})\, , \quad  B^{[3]} \approx -\f{\Psi}{\lm}\f{R}{\sqrt{R^2+3}}  + \mathcal{O}(\lm^{-2/3})\, , \nn \\
 \xi_B &\approx -\arctan{\f{\sqrt{3}}{R}} + \mathcal{O}(\lm^{1/3}) \, ,\label{xiBextreme}
\end{align}
where the factor $\lm$ in the denominators comes from the fact that $U^r \rightarrow \mathcal{O}(\lm)$ as $\lm \rightarrow 0$. If $a$ is exactly equal to 1, the ISCO and the horizon coincide, thus the expression of the magnetic field Eq.~\eqref{B} fails, which is not the situation we are considering. However, the magnetic field is still well-defined in the near-horizon region. From Eq.~\eqref{xiBextreme}, for $a\rightarrow 1$, $\xi_B$ behaves as $d\xi_B/da \sim (1-a^2)^{-5/6}$, and the minimum value takes $\xi_B = -60^{\circ}$ at $r = r_H$, as is demonstrated in the right panel of Fig.~\ref{xiB}.

By taking another coordinate transformation $r = \lm^{2/3} \td{R} +1$, we recover the near-ISCO region, where the radii has the same scaling power with $r_{ms}$ ($\td{R}  = 2^{1/3}$). As the fluid plunges from $r = r_{ms}$, we restrict ourselves to $2^{1/3} > \td{R} \ge 0$. In such region, the magnetic field components are expanded as
\begin{align}
B^{[1]} &\approx \f{\Psi}{\lm^{2/3}} \f{\sqrt{3}}{\td{R}} + \mathcal{O}(1)\, , \quad  B^{[3]} \approx -\f{\Psi}{\lm}\f{\sqrt{2}}{(2^{1/3}-\td{R})^{3/2}}  + \mathcal{O}(\lm^{-1/3})\, ,  \nn  \\
\xi_B &\approx -\arctan{\f{\sqrt{3}(2^{1/3}-\td{R})^{3/2}\lm^{1/3}}{\sqrt{2}\td{R}}} + \mathcal{O}(\lm^{2/3})\, .
\end{align}
The component $B^{[1]}, B^{[3]}$ have different scaling behaviors, and the magnetic field is dominated by $B^{[3]}$ in such region. 

\subsection{The energy extraction efficiency}\label{Ee}

\begin{figure}[h!]
	\centering
	{\includegraphics[width=0.422\textwidth]{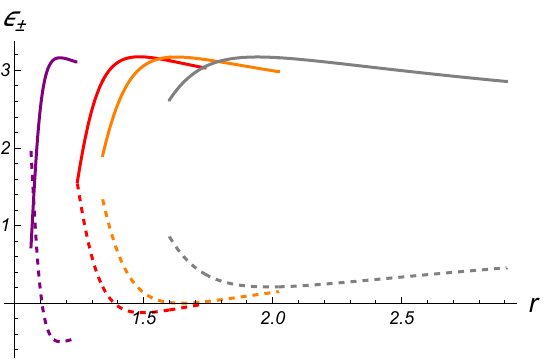}} 
	{\includegraphics[width=0.56\textwidth]{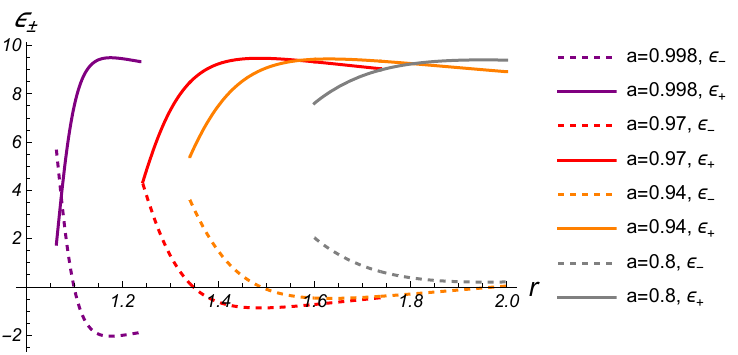}} 
	\caption{The energy-at-infinity per unit enthalpy of the ejected plasmoids as a function of the reconnection location, for differing values of $a$. Given the assumption of a hot plasma, we fix $\td{p} = 1/4$. The magnetization is set to be $\s_0=10$ (left) and $\s_0=100$ (right). }
	\label{epins}
\end{figure}

In this section, we discuss the behavior of the energy-at-infinity per enthalpy of the ejected plasmoids in the plunging plasma, as described by Eq.~\eqref{ep}.  In Fig.~\ref{epins}, we illustrate the variations in $\epsilon_{\pm}$ with different reconnection location.\footnote{Typically, the bulk plasma magnetization $\s_0$ varies with $r$, which involves the details of the thermodynamic distributions of the fluid. For the sake of simplicity in this study, we treat $\s_0$ as a constant across the fluid.} In all cases, $\ep_+$ exhibits a gradual increase first as the $X$-point moves  away from the ISCO, and then a rapid drop as the $X$-point moves towards the event horizon. The peaked radial position of $\ep_+$, located between the horizon and the ISCO, gets larger with an increase of the black hole's spin. On the contrary, $\ep_-$ initially experiences a slight decrease, then a swift rise as the $X$-point approaches the horizon. The radius of the minimum position of $\ep_-$ decreases with a higher black hole spin $a$. It is evident that the energy extraction from the black hole through the reconnection process is possible only when the minimum value of $\ep_-$ is negative. For highly magnetized plasma with $\sigma_0 = 100$, the critical value of the spin $a_c \approx 0.862$, where the minimal value of $\ep_-$ is zero. Energy extraction within the plunging plasma only occurs when $a > a_c$.  The critical value $a_c$ increases with the lowering of the magnetization, such as $a_c \approx 0.887$ for $\sigma_0 = 30$ and $a_c \approx 0.938$ for $\sigma_0 = 10$.

To elucidate the efficiency of energy extraction, we define the local energy extraction efficiency as
\be\label{etad}
\eta = \f{\ep_{+}}{\ep_{+}+\ep_{-}} \, .
\ee
This definition aligns with the convention in \cite{Comisso:2020ykg}. For negative $\ep_-$, the value of $\eta$ consistently exceeds 1. It should be noted that in this definition, we have assumed that only $\ep_-$ can have negative values, a condition that is satisfied in the plunging plasma.

\begin{figure}[h!]
	\centering
	{\includegraphics[width=0.48\textwidth]{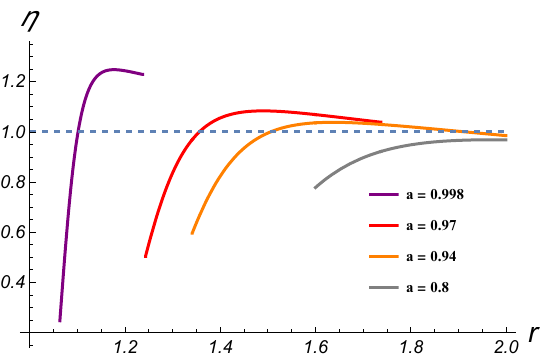}}\, \\
	{\includegraphics[width=0.48\textwidth]{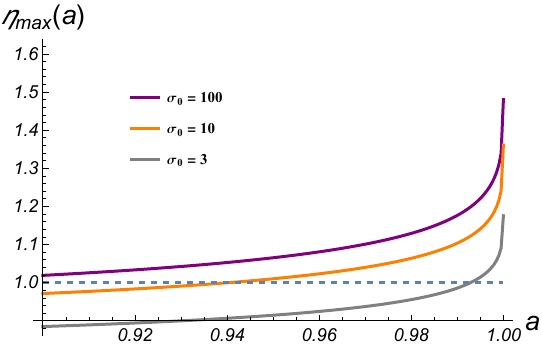}} \,
	{\includegraphics[width=0.48\textwidth]{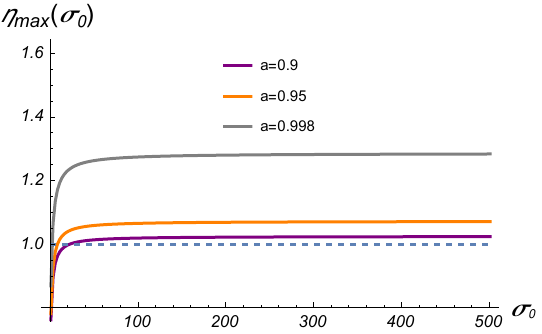}}
	\caption{The local energy extraction efficiency within the plunging plasma. Top: $\eta$ as a function of the reconnection location, for different values of $a$. The magnetization is fixed at $\sigma_0=30$. Bottom left: the maximal value of $\eta$ as a function of the black hole spin. Bottom right: The maximal value of $\eta$ as a function of the magnetization.}
	\label{eta}
\end{figure}

In the top panel of Fig.~\ref{eta}, we depict the behavior of the local energy extraction efficiency as a function of the reconnection location within the ISCO. Our findings indicate that for a strong magnetization scenario where the ISCO is situated inside the ergosphere, i.e., $r_{ms} \leq 2$, the energy extraction is inevitably localized near the ISCO. This observation becomes evident when we consider the limit $\s_0 \gg 1$, which yields $v_{out} \approx 1$, $\g_{out} \approx \sqrt{(1-{\sf g}^2)^3\s_0}/(1+{\sf g}^2)$ with $\sf g$ $= 0.49$ for the high magnetization limit. In the vicinity of the ISCO, we have $U^r, \hat{v}^{(r)}_s  \rightarrow 0$. Hence, from Eq.~\eqref{ep}, Eq.~\eqref{etad} we can derive
\bea\label{iscoeta}
\ep_{\pm} \approx \a \hat{\g}_s \g_{out}  \bigg[ 1+\b^{\phi}\hat{v}_s \pm (\hat{v}_s + \b^{\phi}) \bigg]  \,, \quad \eta \approx 1+  \f{(\b^{\p}-1)(1-\hat{v}_s)}{2(1+\b^{\phi}\hat{v}_s)} \, . 
\eea 
For the case of $a < 0.943$, $r_{ms} < 2$, we observe that $\b^{\phi} < 1$, and consequently, the efficiency remains consistently less than 1. Conversely, for the case where $a > 0.943$, $r_{ms} < 2$,  we have $\b^{\phi} > 1$,  and $\eta > 1$, suggesting that energy extraction is necessarily localized near the ISCO. 

The bottom left panel of Fig.~\ref{eta} displays the maximum value of $\eta$ under various magnetizations, which depends on the spin of the black hole. For moderate magnetization, such as $\sigma_0 = 3$, only black holes with a rather high spin permit energy extraction. In the case of extreme magnetization, exemplified by $\sigma_0 = 100$, the local energy extraction efficiency exceeds 1 across the range of $a \gtrsim 0.9$. In all these scenarios, the rate of increase of $\eta_{max}$ with respect to $a$ escalates considerably as $a$ approaches its extreme value. To estimate the maximum value of $\eta_{max}$ for a near-extreme black hole, we focus on the near-ISCO region, where the fluid velocity behaves as
\bea
\hat{U}_s^{(r)} \approx -\sqrt{\f{2}{3}} \f{(2^{1/3}-\td{R})^{3/2}\lm^{1/3}}{\td{R}} + \mathcal{O}(\lm) \, , \quad  \hat{U}^{(\p)}_s \approx  \f{1}{\sqrt{3}} +  \mathcal{O}(\lm^{2/3})\,, \quad  \hat{v}_s \approx \f{1}{2} +\mathcal{O}(\lm^{2/3}) \, . \label{nearISCOU}
\eea
Combining Eq.~\eqref{nearISCOU} with Eq.~\eqref{iscoeta}, we obtain  
\bea
\ep_{\pm} \approx 0.534 \sqrt{\f{\s_0}{3}} (1\pm 2)\,, \quad \eta_{max} \approx \f{3}{2} \, .
\eea
The maximal value of $\eta$ aligns with those found in \cite{Comisso:2020ykg}. This suggests that despite the fact that the plasma from the ISCO ultimately plunges into the horizon, the fluid flow in the near-ISCO region is nearly circular. Correspondingly, the energy extraction efficiency is peaked in this region for high-spin black holes. For $a\rightarrow 1$, $\eta_{max}$ behaves as $d\eta_{max}/da \sim (1-a^2)^{-5/6}$, a distinctive feature for near-extreme Kerr black holes.

Moreover, in the plunging region, the plasma magnetization can be extremely high, surpassing $\sigma_0=100$ by several orders of magnitude. For this reason, we present the maximum value of $\eta$ as it scales with magnetization in the bottom right panel of  Fig.~\ref{eta}.
We observe that once $\sigma_0$ exceeds approximately 100, the efficiency becomes nearly a constant. Therefore, at least in the study of energy extraction, we can safely use a typical value of $\sigma_0 = 100$ to represent the cases of extremely high magnetization. 

Additionally, we also discuss the case involving the infalling bulk plasma in \ref{infalling}, where we confirm that energy extraction cannot occur within the infalling bulk plasma. 

\subsection{The escaping conditions}\label{Ec}

Even though a negative energy plasmoid ($\ep_-<0$) can be produced, it is essential for the other plasmoid with $\ep_+$ to escape to the infinity, in order for the energy extraction to occur. From an astrophysical perspective, the thermal radiation from the escaping plasmoid, when it is far from the black hole, offers better detectability compared to that within the accretion region. The latter tends to be obscured by background radiation and the in-medium effects. Therefore, in this section, we will scrutinize the escaping conditions for the ejected plasmoid.


For strong magnetization, the plasmoid possesses a relativistic ejection velocity and is nearly collisionless within the bulk plasma after leaving the $X$-point \cite{Priest1986, Titov1997}. Meanwhile, as the thermal energy in the plasmoid cannot convert into the kinetic energy and will be radiated away, it does not affect the plasmoid's trajectory. As a result, we can decouple the plasmoid from the complicated environment after magnetic reconnection, and treat it as a point particle that follows a timelike geodesic. Therefore, we may study the plasmoid's kinetic energy, dictated by the ejection velocity Eq.~\eqref{4v}, to establish the escaping conditions. The geodesic equation in the equatorial plane takes the form
\bea\label{potential}
-\f{1}{2}\left( \f{d r}{d \tau} \right)^2 + \f{\mathcal{E}^2-1}{2} = V(r,  \mathcal{E} , \mathcal{L} ) = -\f{1}{r}+\f{\mathcal{L}^2-a^2(\mathcal{E}^2-1)}{2r^2}-\f{(\mathcal{L}-a\mathcal{E})^2}{r^3} \, ,
\eea	
where $\mathcal{E}$ and $\mathcal{L}$ are the conserved energy density and angular momentum density, respectively, and $V(r,  \mathcal{E} , \mathcal{L} ) $ serves as an effective potential. Generally, for a fixed $\mathcal{E}$, with $\mathcal{L}$ incrementally increasing from zero, an extreme point in the potential will emerge after $\mathcal{L}$ reaching a certain critical value. As $\mathcal{L}$ continues to rise, this extreme point will shift towards a larger radius. 

\begin{figure}[h!]
	\centering
	{\includegraphics[scale=1.1]{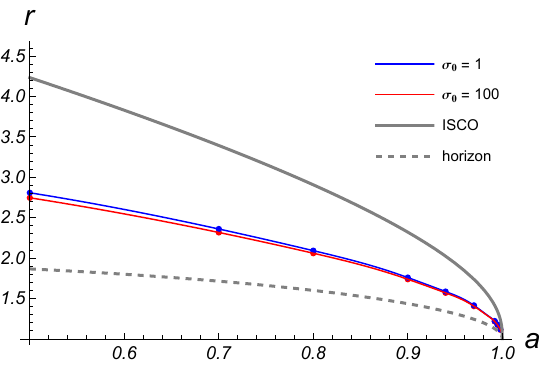}}
	\caption{The critical radius as a function of black hole spin. The magnetization is set to be $\sigma_0 = 1$ (blue curve) and $\sigma_0 = 100$ (red curve). Energy extraction becomes effective only when the $X$-point is situated beyond the critical radius, enabling the plasmoid to escape to infinity. For comparison, we have plotted the ISCO and the event horizon.}
	\label{escape1}
\end{figure}

For the plasmoid, the energy and angular momentum per rest mass are determined by the plasmoid's four-velocity at the $X$-point, as outlined in Eq.~\eqref{4v}. For constant $\sigma_0$ and $\sf g$ (hence a constant $ v_{out}$), we infer from Eq.~\eqref{angle}, Eq.~\eqref{iscoEL} and Eq.~\eqref{iscour} that $\mathcal{E}, \mathcal{L}$ adopt the forms of
\bea
\mathcal{E} = \mathcal{E} \left(\, r_0,\, \xi_B(r_0) \,\right) \, , \quad  \mathcal{L} = \mathcal{L} \left(\, r_0,\, \xi_B(r_0) \,\right) \, ,
\eea
where $r_0$ denotes the radius of the $X$-point.\footnote{Note that unlike the first two sections where $r$ denotes the reconnection location, in this section we specifically use $r_0$ to denote the $X$-point's location, and $r$ is used to represent the radius variable of the plasmoid's motion.} From the equations in Sec. \ref{MFA} and Sec. \ref{Ee}, we see that the radial velocity of the plasmoid with $\ep_+$ always points inward. For the plasmoid to escape, the potential within $r = r_0$ must be sufficiently high to reverse the plasmoid's direction, making it outgoing. We define $r_m$ as the extreme point of the potential. In the model under consideration, $r_m$ is determined by the radius of the $X$-point, $r_m = r_m(r_0)$. The escaping condition is then expressed as inequalities for $r_0$,
\bea\label{con}
\f{\mathcal{E}(r_0)^2-1}{2} < 
\left\{
\begin{aligned}
& \,\, V\left(\,r_m(r_0),\,\mathcal{E}(r_0),\,\mathcal{L}(r_0)\,\right)  \, , \quad \, \,  \text{if} \quad  r_H<r_m(r_0) < r_0 \, , \\
& \,\, V\left(\,r_H,\,\mathcal{E}(r_0),\,\mathcal{L}(r_0)\,\right) \,, \quad \, \,  \text{otherwise} \, .
\end{aligned} \right.
\eea	
Moreover, if the extremal point is situated beyond the $X$-point, i.e., $r_m(r_0) > r_0$, the plasmoid cannot be bounced outward and consequently falls into the horizon. These conditions are sufficient for the plasmoid with  $\ep_+$ to reach the infinity. As expected, Eq.~\eqref{con} assists in identifying a critical radius $r = r_c$, which separates the region $\{r_0\, | \, r_0 \in (r_H, r_{ms})\}$ into two distinct sections. One section is the region where the plasmoid will ultimately be captured by the black hole, while the other permits the plasmoid's escaping. The result of the critical radius is presented in Fig.~\ref{escape1}. We have considered both weak and strong magnetizations, represented by the blue and red curves, respectively. When $1<\sigma_0<100$, the critical radius lies in between the two curves. The result indicates that the critical radius is insensitive to the magnetization $\sigma_0$ but is sensitive to the spin of the black hole. 

\begin{figure}[h!]
	\centering
	{\includegraphics[scale=1.1]{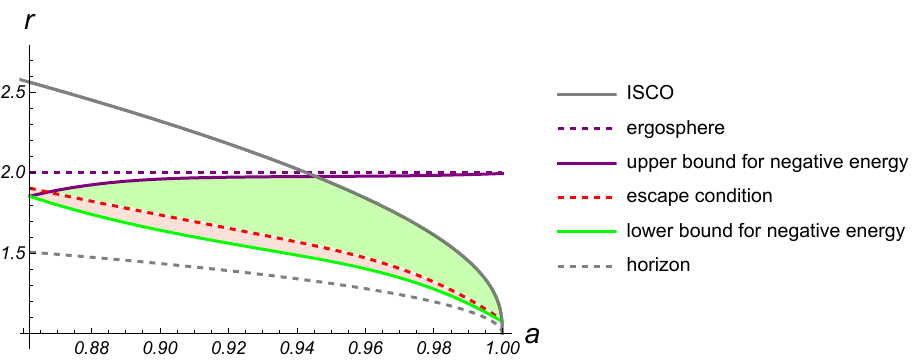}}
	\caption{The region in green permits effective energy extraction via magnetic reconnection. Only between the green and purple curves can $\eta$ be above 1. The red dashed curve stands for the critical radius for the escaping condition, above which the plasmoid with $\ep_+$ can escape to the infinity. The magnetization is set to be $\sigma_0 = 100$. }
	\label{escape2}
\end{figure}

In Fig.~\ref{escape2} we illustrate the allowed region in the radius-spin plane for effective energy extraction in the plunging plasma. The green and purple curves act as the lower and upper bounds for the production of negative energy plasmoids, $\ep_- < 0$. As expected, the upper bound emerges near the ergosphere when $a \simeq 0.862$, and asymptotically  approaches it as $a$ increases. Magnetic reconnection occurring in the region between these two curves results in a local energy extraction efficiency that exceeds one. Furthermore, the red dashed curve, which represents the critical radius for the escaping condition, divides the $\eta >1$ region into two sections, represented in  pink and green respectively. Magnetic reconnection within the pink area produces the plasmoids that plunge into the black hole. On the contrary, magnetic reconnection within the green area can yield a plasmoid that could escape to the infinity, validating the energy extraction process. Fortunately, the escaping restriction still leaves us with a substantial parameter space in green, where energy extraction appears promising.

\section{Summary and Discussions}
\label{sec:sum}
 
In this study, we conducted an in-depth study of energy extraction from a Kerr black hole via magnetic reconnections occurring in the plunging region. We initiated our study by considering a general stationary and axisymmetric spacetime, assuming that the plasma fluid in the equatorial plane adheres to the Killing symmetry of the spacetime. The accompanied magnetic field was set to reverse its direction near the equatorial plane, a region conducive to potent magnetic reconnection. Utilizing the ideal-MHD condition, we obtained the global structure of the magnetic field, which fixes the ejection directions of the plasmoids. Our following analysis incorporated a fast reconnection model characterized by strong magnetic pressure, wherein electromagnetic energy is swiftly liberated, resulting in the formation of two high-velocity plasmoids moving in opposite directions out of the current sheet. Subsequently, we derived the analytical form of the energy-at-infinity per enthalpy for the plasmoids. 

For a Kerr black hole, a plasmoid with negative energy can only be generated within the ergosphere, whose radius exceeds the radius of the ISCO for a moderately spinning black hole, i.e., $a<0.943M$. Therefore, we considered the plasma fluid within the ISCO as the site for magnetic reconnection, with the streamlines following the plunging geodesics from the ISCO to the event horizon.
We illustrated how the energy-at-infinity per enthalpy of the ejected plasmoids varies with the reconnection location under different black hole spin and magnetization parameter. The result revealed a critical spin $a \simeq 0.862M$ for highly magnetized plasma, beyond which energy extraction can occur within the ISCO. With the $X$-point moving towards the horizon, the local energy extraction efficiency initially increases, then rapidly falls below one. The peak value of local efficiency escalates with the black hole spin, exhibiting an infinitely increasing rate as the spin approaches its extreme value.

Furthermore, we investigated the conditions necessary for a plasmoid with positive energy to escape to the infinity, a crucial requirement for the completion of the energy extraction. By scrutinizing the equation of motion for the plasmoid, we pinpointed a critical radius of the $X$-point, beyond which the plasmoid with positive energy can successfully escape. By combining the escaping conditions with the requirement for producing a negative energy plasmoid, we determined the region where energy extraction is effective, as shown in the green region in Fig. \ref{escape2}. Broadly speaking, the lower boundary of the effective region is dictated by the critical radius for the escaping, and the upper boundary is determined by the upper limit for negative energy when the ISCO is situated outside the ergosphere. If the ISCO is located within the ergosphere, it itself serves as the upper boundary of the effective region.

Unlike previous studies that considered magnetic reconnections occurring in bulk plasma following (probably unstable) circular orbits \cite{Koide:2008xr, Comisso:2020ykg}, our research assumed that the fluid plunges from the ISCO to the horizon. In contrast to the toroidal plasma, where the magnetic field cannot be derived from the MHD scheme, the magnetic field in the plunging plasma was determined by the ideal-MHD condition. Thus, the ejection directions of the plasmoids are fixed in our study, rather than being selected phenomenologically. We also analyzed the near-horizon behaviors of the magnetic field and energy extraction efficiency in the high-spin case. To our belief, it would be demanding to systematically investigate the magnetic reconnection model in the near-horizon geometry of a near-extreme Kerr black hole \cite{Bardeen:1999px}, which may differ significantly from those in non-extreme Kerr spacetime.

Building upon the current study, several potential research directions emerge. First, given that the framework in this work is applicable to general stationary and axisymmetric spacetimes, it would be feasible to extend the study in this work to other Kerr-like black holes,  including the rotating black holes in alternative gravity theories and in an expanding universe. Second, our study confines magnetic reconnections to the black hole's equatorial plane. However, numerical simulations suggest that magnetic reconnection can occur near the relativistic jet as well \cite{Yang:2024kpz}, characterized by a parabolic profile beyond the equatorial plane. Therefore, it could be important to investigate the energy extraction near the jet via magnetic reconnection.

Considering that a plasmoid might be significantly accelerated at the $X$-point and escape to the distance, its emission profile would be markedly different from that of the background plasma fluid. Thus, from an astronomical viewpoint, it is meaningful to investigate the optical appearance of such plasmoid \cite{Ripperda:2020bpz, Aimar:2023kzj}. As the observational precision of the Event Horizon Telescope (EHT) and the GRAVITY collaboration continues to improve \cite{EventHorizonTelescope:2024vii, GRAVITY:2018det}, it is anticipated that the characteristics of these plasmoids can be detected in the radio and near-infrared bands in the future \cite{Emami:2022ydq}. This would enable us to identify their energies and origins, and potentially to test energy extraction mechanisms.

\section*{Acknowledgments}

We would like to thank Z.Y. Zhang for  valuable discussions. The work is partly supported by NSFC Grant No. 12275004.
	
\appendix	

\section{Infalling Plasma}\label{infalling}
\begin{figure}[h!]
	\centering
	{\includegraphics[width=0.8\textwidth]{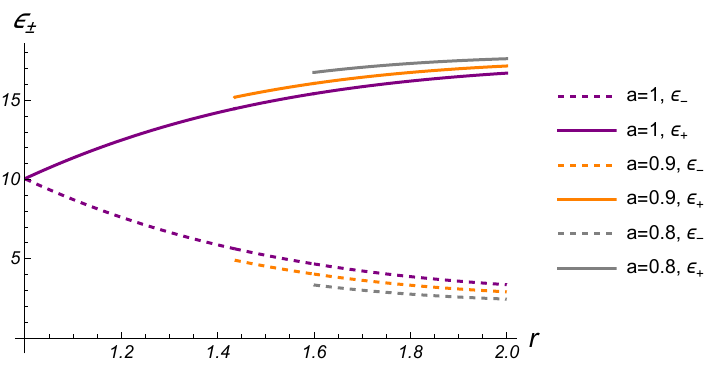}}
	\caption{The energy-at-infinity per enthalpy of the ejected plasmoids as a function of the reconnection location, within the infalling plasma. Assuming a hot plasma, we fix $\td{p} = 1/4$. The magnetization is set to be $\s_0=100$.}
	\label{epinfall}
\end{figure}
In this appendix, we give a brief discussion about the ejection energies of the plasmoids in the bulk plasma freely infalling from the infinity in the Kerr spacetime. The four-velocity of the streamline takes
\bea\label{infalling}
U^r = -\f{\sqrt{2r(r^2+a^2)} }{r^2}\, , \quad  E = 1\, , \quad L = 0 \, .
\eea	
For a large enough $\s_0$ we can drop the last term in Eq.~\eqref{ep}, resulting in
\bea
\ep_{\pm} = \g_{out} \bigg[ 1 \pm v_{out} \big( \hat{v}_s\cos{\xi_B} + \a\b^{\p}\sin{\xi_B}  \big)\bigg]\, ,
\eea
where $\hat{v}_s = r^2(U^r)^2/\sqrt{(r^2+a^2)^2-a^2\D}$, with $\D = r^2-2r+a^2$. Noting that for the infalling plasma, $\a\b^{\p}/ \hat{v}_s = - \tan{\xi_B}$, we find
\be
\ep_{\pm} = \g_{out} \bigg[ 1 \pm v_{out} \sqrt{\f{4a^2 + 2r(r^2+a^2)}{\D r^2+ 2r(r^2+a^2)}} \cos{(2 \xi_B)} \bigg]\, .
\ee
It can be checked that $\ep_+ \geq \ep_-  > 0$, thus there are no negative energy plasmoid after magnetic reconnections, even with a strong magnetization. The behaviors of $\ep_{\pm}$ with  different spin are presented in Fig.~\ref{epinfall}. It is evident that the black hole energy can not be extracted via magnetic reconnection in the infalling plasma.

\bibliographystyle{utphys}
\bibliography{references}

\providecommand{\href}[2]{#2}\begingroup\raggedright\begin{thebibliography}{10}

\bibitem{Christodoulou:1970wf}
D.~Christodoulou, ``{Reversible and irreversible transforations in black hole
  physics},'' \href{http://dx.doi.org/10.1103/PhysRevLett.25.1596}{{\em Phys.
  Rev. Lett.} {\bfseries 25} (1970) 1596--1597}.

\bibitem{Wald:1974kya}
R.~M. Wald, ``{Energy Limits on the Penrose Process},''
  \href{http://dx.doi.org/10.1086/152959}{{\em Astrophys. J.} {\bfseries 191}
  (1974) 231}.

\bibitem{Brito:2015oca}
R.~Brito, V.~Cardoso, and P.~Pani, ``{Superradiance}: {New Frontiers in Black
  Hole Physics},'' \href{http://dx.doi.org/10.1007/978-3-319-19000-6}{{\em
  Lect. Notes Phys.} {\bfseries 906} (2015) pp.1--237},
  \href{http://arxiv.org/abs/1501.06570}{{\ttfamily arXiv:1501.06570 [gr-qc]}}.

\bibitem{Blandford:1977ds}
R.~D. Blandford and R.~L. Znajek, ``{Electromagnetic extractions of energy from
  Kerr black holes},'' \href{http://dx.doi.org/10.1093/mnras/179.3.433}{{\em
  Mon. Not. Roy. Astron. Soc.} {\bfseries 179} (1977) 433--456}.

\bibitem{Komissarov:2005wj}
S.~S. Komissarov, ``{Observations of the Blandford-Znajek and the MHD Penrose
  processes in computer simulations of black hole magnetospheres},''
  \href{http://dx.doi.org/10.1111/j.1365-2966.2005.08974.x}{{\em Mon. Not. Roy.
  Astron. Soc.} {\bfseries 359} (2005) 801--808},
  \href{http://arxiv.org/abs/astro-ph/0501599}{{\ttfamily
  arXiv:astro-ph/0501599}}.

\bibitem{Parfrey:2018dnc}
K.~Parfrey, A.~Philippov, and B.~Cerutti, ``{First-Principles Plasma
  Simulations of Black-Hole Jet Launching},''
  \href{http://dx.doi.org/10.1103/PhysRevLett.122.035101}{{\em Phys. Rev.
  Lett.} {\bfseries 122} no.~3, (2019) 035101},
  \href{http://arxiv.org/abs/1810.03613}{{\ttfamily arXiv:1810.03613
  [astro-ph.HE]}}.

\bibitem{Lu:2023bbn}
R.-S. Lu {\em et~al.}, ``{A ring-like accretion structure in M87 connecting its
  black hole and jet},''
  \href{http://dx.doi.org/10.1038/s41586-023-05843-w}{{\em Nature} {\bfseries
  616} no.~7958, (2023) 686--690},
  \href{http://arxiv.org/abs/2304.13252}{{\ttfamily arXiv:2304.13252
  [astro-ph.HE]}}.

\bibitem{yamada2009}
E.~Zweibel and M.~Yamada, ``Magnetic reconnection in astrophysical and
  laboratory plasmas,''
  \href{http://dx.doi.org/10.1146/annurev-astro-082708-101726}{{\em Annual
  Review of Astronomy and Astrophysics - ANNU REV ASTRON ASTROPHYS} {\bfseries
  47} (09, 2009) 291--332}.

\bibitem{SP1}
P.~A. {Sweet}, ``{The Neutral Point Theory of Solar Flares},'' in {\em
  Electromagnetic Phenomena in Cosmical Physics}, B.~{Lehnert}, ed., vol.~6,
  p.~123.
\newblock Jan., 1958.

\bibitem{Petschek}
H.~E. {Petschek}, ``{Magnetic Field Annihilation},'' in {\em NASA Special
  Publication}, vol.~50, p.~425.
\newblock 1964.

\bibitem{rela-SP}
Y.~E. Lyubarsky, ``{On the relativistic magnetic reconnection},''
  \href{http://dx.doi.org/10.1111/j.1365-2966.2005.08767.x}{{\em Mon. Not. Roy.
  Astron. Soc.} {\bfseries 358} (2005) 113--119},
  \href{http://arxiv.org/abs/astro-ph/0501392}{{\ttfamily
  arXiv:astro-ph/0501392}}.

\bibitem{Koide:2008xr}
S.~Koide and K.~Arai, ``{Energy Extraction from a Rotating Black Hole by
  Magnetic Reconnection in Ergosphere},''
  \href{http://dx.doi.org/10.1086/589497}{{\em Astrophys. J.} {\bfseries 682}
  (2008) 1124}, \href{http://arxiv.org/abs/0805.0044}{{\ttfamily
  arXiv:0805.0044 [astro-ph]}}.

\bibitem{Sironi:2022hnw}
L.~Sironi, ``{Nonideal Fields Solve the Injection Problem in Relativistic
  Reconnection},'' \href{http://dx.doi.org/10.1103/PhysRevLett.128.145102}{{\em
  Phys. Rev. Lett.} {\bfseries 128} no.~14, (2022) 145102},
  \href{http://arxiv.org/abs/2203.04342}{{\ttfamily arXiv:2203.04342
  [astro-ph.HE]}}.

\bibitem{French:2022zfv}
O.~French, F.~Guo, Q.~Zhang, and D.~A. Uzdensky, ``{Particle Injection and
  Nonthermal Particle Acceleration in Relativistic Magnetic Reconnection*},''
  \href{http://dx.doi.org/10.3847/1538-4357/acb7dd}{{\em Astrophys. J.}
  {\bfseries 948} no.~1, (2023) 19},
  \href{http://arxiv.org/abs/2210.08358}{{\ttfamily arXiv:2210.08358
  [astro-ph.HE]}}.

\bibitem{Dong:2022crn}
C.~Dong, L.~Wang, Y.-M. Huang, L.~Comisso, T.~A. Sandstrom, and
  A.~Bhattacharjee, ``{Reconnection-driven energy cascade in
  magnetohydrodynamic turbulence},''
  \href{http://dx.doi.org/10.1126/sciadv.abn7627}{{\em Sci. Adv.} {\bfseries 8}
  no.~49, (2022) abn7627}, \href{http://arxiv.org/abs/2210.10736}{{\ttfamily
  arXiv:2210.10736 [astro-ph.SR]}}.

\bibitem{Comisso:2023ygd}
L.~Comisso and B.~Jiang, ``{Pitch-angle Anisotropy Imprinted by Relativistic
  Magnetic Reconnection},''
  \href{http://dx.doi.org/10.3847/1538-4357/ad1241}{{\em Astrophys. J.}
  {\bfseries 959} no.~2, (2023) 137},
  \href{http://arxiv.org/abs/2310.17560}{{\ttfamily arXiv:2310.17560
  [astro-ph.HE]}}.

\bibitem{Comisso:2020ykg}
L.~Comisso and F.~A. Asenjo, ``{Magnetic Reconnection as a Mechanism for Energy
  Extraction from Rotating Black Holes},''
  \href{http://dx.doi.org/10.1103/PhysRevD.103.023014}{{\em Phys. Rev. D}
  {\bfseries 103} no.~2, (2021) 023014},
  \href{http://arxiv.org/abs/2012.00879}{{\ttfamily arXiv:2012.00879
  [astro-ph.HE]}}.

\bibitem{Liu2017}
Y.-H. Liu, M.~Hesse, F.~Guo, W.~Daughton, H.~Li, P.~A. Cassak, and M.~A. Shay,
  ``{Why does steady-state magnetic reconnection have a maximum local rate of
  order 0.1?},'' \href{http://dx.doi.org/10.1103/PhysRevLett.118.085101}{{\em
  Phys. Rev. Lett.} {\bfseries 118} no.~8, (2017) 085101},
  \href{http://arxiv.org/abs/1611.07859}{{\ttfamily arXiv:1611.07859
  [physics.plasm-ph]}}.

\bibitem{Khodadi:2022dff}
M.~Khodadi, ``{Magnetic reconnection and energy extraction from a spinning
  black hole with broken Lorentz symmetry},''
  \href{http://dx.doi.org/10.1103/PhysRevD.105.023025}{{\em Phys. Rev. D}
  {\bfseries 105} no.~2, (2022) 023025},
  \href{http://arxiv.org/abs/2201.02765}{{\ttfamily arXiv:2201.02765 [gr-qc]}}.

\bibitem{Carleo:2022qlv}
A.~Carleo, G.~Lambiase, and L.~Mastrototaro, ``{Energy extraction via magnetic
  reconnection in Lorentz breaking Kerr\textendash{}Sen and Kiselev black
  holes},'' \href{http://dx.doi.org/10.1140/epjc/s10052-022-10751-w}{{\em Eur.
  Phys. J. C} {\bfseries 82} no.~9, (2022) 776},
  \href{http://arxiv.org/abs/2206.12988}{{\ttfamily arXiv:2206.12988 [gr-qc]}}.

\bibitem{Wei:2022jbi}
S.-W. Wei, H.-M. Wang, Y.-P. Zhang, and Y.-X. Liu, ``{Effects of tidal charge
  on magnetic reconnection and energy extraction from spinning braneworld black
  hole},'' \href{http://dx.doi.org/10.1088/1475-7516/2022/04/050}{{\em JCAP}
  {\bfseries 04} no.~04, (2022) 050},
  \href{http://arxiv.org/abs/2201.12729}{{\ttfamily arXiv:2201.12729 [gr-qc]}}.

\bibitem{Liu:2022qnr}
W.~Liu, ``{Energy Extraction via Magnetic Reconnection in the Ergosphere of a
  Rotating Non-Kerr Black Hole},''
  \href{http://dx.doi.org/10.3847/1538-4357/ac3de3}{{\em Astrophys. J.}
  {\bfseries 925} no.~2, (2022) 149},
  \href{http://arxiv.org/abs/2204.07338}{{\ttfamily arXiv:2204.07338
  [astro-ph.HE]}}.

\bibitem{Wang:2022qmg}
C.-H. Wang, C.-Q. Pang, and S.-W. Wei, ``{Extracting energy via magnetic
  reconnection from Kerr\textendash{}de Sitter black holes},''
  \href{http://dx.doi.org/10.1103/PhysRevD.106.124050}{{\em Phys. Rev. D}
  {\bfseries 106} no.~12, (2022) 124050},
  \href{http://arxiv.org/abs/2209.08837}{{\ttfamily arXiv:2209.08837 [gr-qc]}}.

\bibitem{Li:2023nmy}
Z.~Li, X.-K. Guo, and F.~Yuan, ``{Energy extraction from rotating regular black
  hole via Comisso-Asenjo mechanism},''
  \href{http://arxiv.org/abs/2304.08831}{{\ttfamily arXiv:2304.08831 [gr-qc]}}.

\bibitem{Li:2023htz}
Z.~Li and F.~Yuan, ``{Energy extraction via Comisso-Asenjo mechanism from
  rotating hairy black hole},''
  \href{http://arxiv.org/abs/2304.12553}{{\ttfamily arXiv:2304.12553 [gr-qc]}}.

\bibitem{Ye:2023xyv}
X.~Ye, C.-H. Wang, and S.-W. Wei, ``{Extracting spinning wormhole energy via
  magnetic reconnection},'' \href{http://arxiv.org/abs/2306.12097}{{\ttfamily
  arXiv:2306.12097 [gr-qc]}}.

\bibitem{Khodadi:2023juk}
M.~Khodadi, D.~F. Mota, and A.~Sheykhi, ``{Harvesting energy driven by
  Comisso-Asenjo process from Kerr-MOG black holes},''
  \href{http://dx.doi.org/10.1088/1475-7516/2023/10/034}{{\em JCAP} {\bfseries
  10} (2023) 034}, \href{http://arxiv.org/abs/2307.00478}{{\ttfamily
  arXiv:2307.00478 [astro-ph.HE]}}.

\bibitem{Shaymatov:2023dtt}
S.~Shaymatov, M.~Alloqulov, B.~Ahmedov, and A.~Wang, ``{A Kerr-Newman-MOG black
  hole's impact on the magnetic reconnection},''
  \href{http://arxiv.org/abs/2307.03012}{{\ttfamily arXiv:2307.03012 [gr-qc]}}.

\bibitem{Zhang:2024rvk}
S.-J. Zhang, ``{Energy extraction via magnetic reconnection in
  Konoplya-Rezzolla-Zhidenko parametrized black holes},''
  \href{http://dx.doi.org/10.1103/PhysRevD.109.084066}{{\em Phys. Rev. D}
  {\bfseries 109} no.~8, (2024) 084066},
  \href{http://arxiv.org/abs/2402.15050}{{\ttfamily arXiv:2402.15050 [gr-qc]}}.

\bibitem{Mummery:2022ana}
A.~Mummery and S.~Balbus, ``{Inspirals from the Innermost Stable Circular Orbit
  of Kerr Black Holes: Exact Solutions and Universal Radial Flow},''
  \href{http://dx.doi.org/10.1103/PhysRevLett.129.161101}{{\em Phys. Rev.
  Lett.} {\bfseries 129} no.~16, (2022) 161101},
  \href{http://arxiv.org/abs/2209.03579}{{\ttfamily arXiv:2209.03579 [gr-qc]}}.

\bibitem{Liu:2023tcy}
Y.~Liu and B.~Sun, ``{Analytical solutions of equatorial geodesic motion in
  Kerr spacetime*},'' \href{http://dx.doi.org/10.1088/1674-1137/ad260a}{{\em
  Chin. Phys. C} {\bfseries 48} no.~4, (2024) 045107},
  \href{http://arxiv.org/abs/2305.11045}{{\ttfamily arXiv:2305.11045 [gr-qc]}}.

\bibitem{Machida:2002ub}
M.~Machida and R.~Matsumoto, ``{Global three - dimensional MHD simulations of
  black hole accretion disks: X-ray flares in the plunging region},''
  \href{http://dx.doi.org/10.1086/346070}{{\em Astrophys. J.} {\bfseries 585}
  (2003) 429--442}, \href{http://arxiv.org/abs/astro-ph/0211240}{{\ttfamily
  arXiv:astro-ph/0211240}}.

\bibitem{Reynolds:2006uq}
C.~S. Reynolds, D.~Garofalo, and M.~C. Begelman, ``{Trapping of magnetic flux
  by the plunge region of a black hole accretion disk},''
  \href{http://dx.doi.org/10.1086/507691}{{\em Astrophys. J.} {\bfseries 651}
  (2006) 1023--1030}, \href{http://arxiv.org/abs/astro-ph/0607381}{{\ttfamily
  arXiv:astro-ph/0607381}}.

\bibitem{Zhu:2012vf}
Y.~Zhu, S.~W. Davis, R.~Narayan, A.~K. Kulkarni, R.~F. Penna, and J.~E.
  McClintock, ``{The Eye of the Storm: Light from the Inner Plunging Region of
  Black Hole Accretion Discs},''
  \href{http://dx.doi.org/10.1111/j.1365-2966.2012.21181.x}{{\em Mon. Not. Roy.
  Astron. Soc.} {\bfseries 424} (2012) 2504},
  \href{http://arxiv.org/abs/1202.1530}{{\ttfamily arXiv:1202.1530
  [astro-ph.HE]}}.

\bibitem{Wilkins:2020pgu}
D.~R. Wilkins, C.~S. Reynolds, and A.~C. Fabian, ``{Venturing beyond the ISCO:
  Detecting X-ray emission from the plunging regions around black holes},''
  \href{http://dx.doi.org/10.1093/mnras/staa628}{{\em Mon. Not. Roy. Astron.
  Soc.} {\bfseries 493} no.~4, (2020) 5532--5550},
  \href{http://arxiv.org/abs/2003.00019}{{\ttfamily arXiv:2003.00019
  [astro-ph.HE]}}.

\bibitem{Hankla:2022iwk}
A.~M. Hankla, N.~Scepi, and J.~Dexter, ``{Non-thermal emission from the
  plunging region: a model for the high-energy tail of black hole X-ray binary
  soft states},'' \href{http://dx.doi.org/10.1093/mnras/stac1785}{{\em Mon.
  Not. Roy. Astron. Soc.} {\bfseries 515} no.~1, (2022) 775--784},
  \href{http://arxiv.org/abs/2206.12018}{{\ttfamily arXiv:2206.12018
  [astro-ph.HE]}}.

\bibitem{Dong:2023bbd}
J.~Dong, G.~Mastroserio, J.~A. Garc\i{}a, A.~Ingram, E.~Nathan, and R.~Connors,
  ``{X-ray Reflection from the Plunging Region of Black Hole Accretion
  Disks},'' \href{http://arxiv.org/abs/2312.09210}{{\ttfamily arXiv:2312.09210
  [astro-ph.HE]}}.

\bibitem{Ruffini:1975ne}
R.~Ruffini and J.~R. Wilson, ``{Relativistic Magnetohydrodynamical Effects of
  Plasma Accreting Into a Black Hole},''
  \href{http://dx.doi.org/10.1103/PhysRevD.12.2959}{{\em Phys. Rev. D}
  {\bfseries 12} (1975) 2959}.

\bibitem{SP2}
E.~N. {Parker}, ``{Sweet's Mechanism for Merging Magnetic Fields in Conducting
  Fluids},'' \href{http://dx.doi.org/10.1029/JZ062i004p00509}{{\em Journal of
  Geophysical Research} {\bfseries 62} no.~4, (Dec., 1957) 509--520}.

\bibitem{substorm}
E.~N. {Parker}, ``{The Solar-Flare Phenomenon and the Theory of Reconnection
  and Annihiliation of Magnetic Fields.},''
  \href{http://dx.doi.org/10.1086/190087}{{\em Astrophys. J.} {\bfseries 8}
  (July, 1963) 177}.

\bibitem{anti-Petschek1}
D.~{Biskamp}, ``{Magnetic reconnection via current sheets},''
  \href{http://dx.doi.org/10.1063/1.865670}{{\em Physics of Fluids} {\bfseries
  29} no.~5, (May, 1986) 1520--1531}.

\bibitem{anti-Petschek2}
T.~{Sato} and T.~{Hayashi}, ``{Externally driven magnetic reconnection and a
  powerful magnetic energy converter},''
  \href{http://dx.doi.org/10.1063/1.862721}{{\em Physics of Fluids} {\bfseries
  22} no.~6, (June, 1979) 1189--1202}.

\bibitem{MAD}
R.~Narayan, I.~V. Igumenshchev, and M.~A. Abramowicz, ``{Magnetically arrested
  disk: an energetically efficient accretion flow},''
  \href{http://dx.doi.org/10.1093/pasj/55.6.L69}{{\em Publ. Astron. Soc. Jap.}
  {\bfseries 55} (2003) L69},
  \href{http://arxiv.org/abs/astro-ph/0305029}{{\ttfamily
  arXiv:astro-ph/0305029}}.

\bibitem{Tchekho2011}
A.~{Tchekhovskoy}, R.~{Narayan}, and J.~C. {McKinney}, ``{Efficient generation
  of jets from magnetically arrested accretion on a rapidly spinning black
  hole},'' \href{http://dx.doi.org/10.1111/j.1745-3933.2011.01147.x}{{\em Mon.
  Not. Roy. Astron. Soc.} {\bfseries 418} no.~1, (Nov., 2011) L79--L83},
  \href{http://arxiv.org/abs/1108.0412}{{\ttfamily arXiv:1108.0412
  [astro-ph.HE]}}.

\bibitem{Priest1986}
E.~R. Priest and T.~G. Forbes, ``New models for fast steady state magnetic
  reconnection,''
  \href{http://dx.doi.org/https://doi.org/10.1029/JA091iA05p05579}{{\em Journal
  of Geophysical Research: Space Physics} {\bfseries 91} no.~A5, (1986)
  5579--5588},
  \href{http://arxiv.org/abs/https://agupubs.onlinelibrary.wiley.com/doi/pdf/10.1029/JA091iA05p05579}{{\ttfamily
  https://agupubs.onlinelibrary.wiley.com/doi/pdf/10.1029/JA091iA05p05579}}.
  \url{https://agupubs.onlinelibrary.wiley.com/doi/abs/10.1029/JA091iA05p05579}.

\bibitem{Titov1997}
V.~Titov and E.~Priest, ``Linear theory of steady x-point magnetic
  reconnection,''
  \href{http://dx.doi.org/https://doi.org/10.1016/S0273-1177(97)00074-4}{{\em
  Advances in Space Research} {\bfseries 19} no.~12, (1997) 1777--1780}.
  \url{https://www.sciencedirect.com/science/article/pii/S0273117797000744}.
  Proceedings of the D0.4 Symposium of COSPAR Scientific Commission D.

\bibitem{Hou:2023bep}
Y.~Hou, Z.~Zhang, M.~Guo, and B.~Chen, ``{A new analytical model of
  magnetofluids surrounding rotating black holes},''
  \href{http://dx.doi.org/10.1088/1475-7516/2024/02/030}{{\em JCAP} {\bfseries
  02} (2024) 030}, \href{http://arxiv.org/abs/2309.13304}{{\ttfamily
  arXiv:2309.13304 [gr-qc]}}.

\bibitem{chandrasekhar1998mathematical}
S.~Chandrasekhar, {\em The Mathematical Theory of Black Holes}.
\newblock International series of monographs on physics. Clarendon Press, 1998.
\newblock \url{https://books.google.co.jp/books?id=LBOVcrzFfhsC}.

\bibitem{Bardeen:1999px}
J.~M. Bardeen and G.~T. Horowitz, ``{The Extreme Kerr throat geometry: A Vacuum
  analog of AdS(2) x S**2},''
  \href{http://dx.doi.org/10.1103/PhysRevD.60.104030}{{\em Phys. Rev. D}
  {\bfseries 60} (1999) 104030},
  \href{http://arxiv.org/abs/hep-th/9905099}{{\ttfamily arXiv:hep-th/9905099}}.

\bibitem{Yang:2024kpz}
H.~Yang, F.~Yuan, H.~Li, Y.~Mizuno, F.~Guo, R.~Lu, L.~C. Ho, X.~Lin, A.~A.
  Zdziarski, and J.~Wang, ``{Modeling the inner part of the jet in M87:
  Confronting jet morphology with theory},''
  \href{http://dx.doi.org/10.1126/sciadv.adn3544}{{\em Sci. Adv.} {\bfseries
  10} no.~12, (2024) adn3544},
  \href{http://arxiv.org/abs/2403.15950}{{\ttfamily arXiv:2403.15950
  [astro-ph.HE]}}.

\bibitem{Ripperda:2020bpz}
B.~Ripperda, F.~Bacchini, and A.~Philippov, ``{Magnetic Reconnection and Hot
  Spot Formation in Black Hole Accretion Disks},''
  \href{http://dx.doi.org/10.3847/1538-4357/ababab}{{\em Astrophys. J.}
  {\bfseries 900} no.~2, (2020) 100},
  \href{http://arxiv.org/abs/2003.04330}{{\ttfamily arXiv:2003.04330
  [astro-ph.HE]}}.

\bibitem{Aimar:2023kzj}
N.~Aimar, A.~Dmytriiev, F.~H. Vincent, I.~E. Mellah, T.~Paumard, G.~Perrin, and
  A.~Zech, ``{Magnetic reconnection plasmoid model for Sagittarius A*
  flares},'' \href{http://dx.doi.org/10.1051/0004-6361/202244936}{{\em Astron.
  Astrophys.} {\bfseries 672} (2023) A62},
  \href{http://arxiv.org/abs/2301.11874}{{\ttfamily arXiv:2301.11874
  [astro-ph.HE]}}.

\bibitem{EventHorizonTelescope:2024vii}
{\bfseries Event Horizon Telescope} Collaboration, K.~Akiyama {\em et~al.},
  ``{First Sagittarius A* Event Horizon Telescope Results. VII. Polarization of
  the Ring},'' \href{http://dx.doi.org/10.3847/2041-8213/ad2df0}{{\em
  Astrophys. J. Lett.} {\bfseries 964} no.~2, (2024) L25}.

\bibitem{GRAVITY:2018det}
{\bfseries GRAVITY} Collaboration, R.~Abuter {\em et~al.}, ``{Detection of
  orbital motions near the last stable circular orbit of the massive black hole
  SgrA},'' \href{http://dx.doi.org/10.1051/0004-6361/201834294}{{\em Astron.
  Astrophys.} {\bfseries 618} (2018) L10},
  \href{http://arxiv.org/abs/1810.12641}{{\ttfamily arXiv:1810.12641
  [astro-ph.GA]}}.

\bibitem{Emami:2022ydq}
R.~Emami {\em et~al.}, ``{Tracing Hot Spot Motion in Sagittarius A* Using the
  Next-Generation Event Horizon Telescope (ngEHT)},''
  \href{http://dx.doi.org/10.3390/galaxies11010023}{{\em Galaxies} {\bfseries
  11} no.~1, (2023) 23}, \href{http://arxiv.org/abs/2211.06773}{{\ttfamily
  arXiv:2211.06773 [astro-ph.GA]}}.

\end{thebibliography}\endgroup
		
\end{document}